\documentclass[journal]{IEEEtran}

\ifCLASSINFOpdf

\else

\fi

\usepackage{mathrsfs}
\usepackage{amsmath}
\usepackage{amsfonts}
\usepackage{amsthm}
\usepackage{amssymb}
\usepackage{graphicx}
\usepackage{subfigure}
\usepackage{indentfirst}
\usepackage{array}
\usepackage{epstopdf}
\usepackage{cite}
\usepackage{enumerate}
\usepackage{bm}
\usepackage{dsfont}
\usepackage[linesnumbered,ruled,vlined]{algorithm2e}
\usepackage{multirow}
\usepackage{verbatim}
\usepackage{stfloats}
\usepackage{makecell}
\usepackage{cancel}
\usepackage{threeparttable}
\usepackage[dvipsnames]{xcolor}
\usepackage[colorlinks=true, linkcolor=blue, citecolor=blue]{hyperref}

\SetKwComment{Comment}{//}{}

\begin{document}

\title{{Deep Generative Modeling Reshapes Compression and Transmission: From Efficiency to Resiliency}}

\author{Jincheng~Dai,
	    Xiaoqi~Qin,
        Sixian~Wang,
        Lexi Xu,
        Kai~Niu,
        and Ping~Zhang

\thanks{This work was supported in part by the National Natural Science Foundation of China under Grant 62293481, Grant 62371063, Grant 92267301, in part by the Beijing Natural Science Foundation under Grant L232047, Grant 4222012.}

\thanks{Jincheng Dai, Xiaoqi Qin, Sixian Wang, Kai Niu and Ping Zhang are with Beijing University of Posts and Telecommunications, Beijing, China.}

\thanks{Lexi Xu is with the Research Institute, China United Network Communications Corporation, Beijing, China.}

\vspace{-1em}
}

\maketitle

\begin{abstract}
Information theory and machine learning are inextricably linked and have even been referred to as ``two sides of the same coin''. One particularly elegant connection is the essential equivalence between probabilistic generative modeling and data compression or transmission. In this article, we reveal the dual-functionality of deep generative models that reshapes both data compression for efficiency and transmission error concealment for resiliency. We present how the contextual predictive capabilities of powerful generative models can be well positioned to be strong compressors and estimators. In this sense, we advocate for viewing the deep generative modeling problem through the lens of end-to-end communications, and evaluate the compression and error restoration capabilities of foundation generative models. We show that the kernel of many large generative models is powerful predictor that can capture complex relationships among semantic latent variables, and the communication viewpoints provide novel insights into semantic feature tokenization, contextual learning, and usage of deep generative models. In summary, our article highlights the essential connections of generative AI to source and channel coding techniques, and motivates researchers to make further explorations in this emerging topic.
	
\end{abstract}

\section{Introduction}\label{section_introduction}

Information theory and machine learning are interconnected deeply and have even been referred to as ``two sides of the same coin'' \cite{mackay2003information}. Specifically, Shannon information theory bifurcates into data compression, optimizing for efficiency, and distortion correction, responsible for reliability.  Deep learning models also have two categories: discriminative models, which specialize in instance fitting, and generative models, dedicated to distribution fitting. In this article, we reveal a particularly profound connection: deep probabilistic generative modeling aligns remarkably with both pivotal branches of information theory. This alignment not only enriches our comprehension of these models but also paves the way for innovative applications in enhancing compression efficiency and transmission robustness. Such insights are pivotal in reshaping the landscape of end-to-end communication systems in the era of generative artificial intelligence (GAI).

Two critical insights have pushed large-scale neural networks revolutionizing generative modeling over the last few years, enabling them with unprecedented ability to capture complex relationships among numerous variables. The first insight to the processing efficiency of deep generative models is \emph{departure to latent space}, that applies auto-encoder-like \emph{nonlinear transform} to the raw data space to find a perceptually equivalent yet computationally more suitable latent space \cite{rombach2022high,balle2020nonlinear}, in which we can efficiently train generative models for large-size data synthesis. The second insight, crucial for the expressive power of mainstream deep generative models -- including auto-regressive models, flow-based models, deep variational auto-encoders (VAEs), and diffusion models -- is the decomposition of learned joint distribution into a sequence of correlated steps \cite{jakub2022deep}. This divide-and-conquer strategy can effectively circumvent the ``curse of dimensionality'', a significant challenge that arises when attempting to explicitly define interactions among a vast array of latent variables. Incidentally, these latent space generative models are consistently trained with the loss function minimized by the negative log-probability over all of the latents, that equals to the expected message length of bits of an optimal entropy codec to describe latent variables \cite{deletang2023language}. There is therefore a direct correspondence between learning a latent generative model with maximum likelihood and training it for data compression with bit length as short as possible.

In the aspect of data transmission utilizing deep generative models, a paramount focus is placed on robustness and error-resilience capabilities. Recent advances in learning-based joint source-channel coding have sparked significant innovations in this field \cite{gunduz2022beyond}. Within the traditional architecture of separate source compression and channel transmission, channel decoding left errors mirroring on loss latents can be effectively mitigated. This is achieved by integrating BERT-like bidirectional Transformers \cite{chang2022maskgit}, which are adept at predicting randomly masked tokens via contextual learning. Alternatively, in a joint source-channel coding framework \cite{bourtsoulatze2019deep}, the natural analog signal produced by variational learned codecs \cite{dai2022nonlinear} exhibit robust performance against impairments typical in imperfect wireless channels. These generative models, operating in the semantic latent space, progressively refine source data through generating latent features. This refinement is conditioned iteratively on the outputs of previous generations that sufficiently leverages the contextual correlations of latents, showcasing an effective and efficient approach to distortion restoration. These novel insights and techniques signify a paradigm shift in our approach to error concealment, and mark new avenues in the application and understanding of deep generative models for achieving resilience in channel transmission.

In this work, we advocate for a novel viewpoint to understanding foundation generative models, framing them within the context of compression and transmission processes. We conceptualize the distribution fitting challenge inherent in deep generative models as a communicative exchange between two entities: a sender Alice, who has access to some data, and Bob, who aims to receive this data using the minimal amount of bits. In this scenario, Alice transmits a series of messages to Bob, each revealing some aspect of the data. Bob's role is to predict these messages, where more accurate predictions lead to reduced bit requirements for transmission. Upon receiving each message, Bob refines his predictions for subsequent messages, collectively forming a comprehensive representation of the data that Alice has just accessed. The efficiency of such a communication process is quantified by a loss function (for model training), defined as the total number of bits required for the entire message transmission.  This framework illustrates the dual-functionality of deep probabilistic generative models: Alice employs them to transform each message as bit sequences (entropy coding), while Bob utilizes them for the prediction of latent variables yet to be received (generation). This dual role underscores the contribution of generative models to both efficiency and resiliency aspects of end-to-end communication systems. Our exploration in this article highlights the essential connections of generative AI to source and channel coding techniques, and motivates further research into this burgeoning field.

\section{Deep Generative Modeling for Compression: The Functionality of Efficiency}\label{section_compression}

The emerging field of compression algorithms using deep neural networks is called \emph{neural compression}. Neural compression becomes a leading trend in developing new codecs where neural networks replace parts of the standard codecs, or entirely neural codecs are trained together with quantization and entropy coding. We will not discuss the general compression schemes in detail but here it is important to understand why deep generative modeling is important in the context of neural compression. The answer was given a long time ago by Shannon who showed in that (informally):
 
\emph{The length of a compressed message representing a source data is proportional to the entropy of this data.}

However, for the end of data compression, the true entropy of data remains elusive due to the unknown nature of the underlying probability distribution, but deep generative models have emerged as an effective tool for estimating this distribution, sparking a growing interest in leveraging them to enhance neural compression techniques \cite{balle2020nonlinear}. This represents \emph{a paradigm shift in probability estimation, transitioning from the traditional statistically counting manner to the modern data-driven learning manner}. These generative models offer dual-advantages: they serve not only as sophisticated tools for modeling the probability distributions necessary for entropy coding but also as a means to substantially enhance the quality of data reconstruction and compression. It is achieved through the integration of innovative generation schemes.

\begin{figure*}[t]
	\setlength{\abovecaptionskip}{0.cm}
	\setlength{\belowcaptionskip}{-0.cm}
	\centering{\includegraphics[scale=0.48]{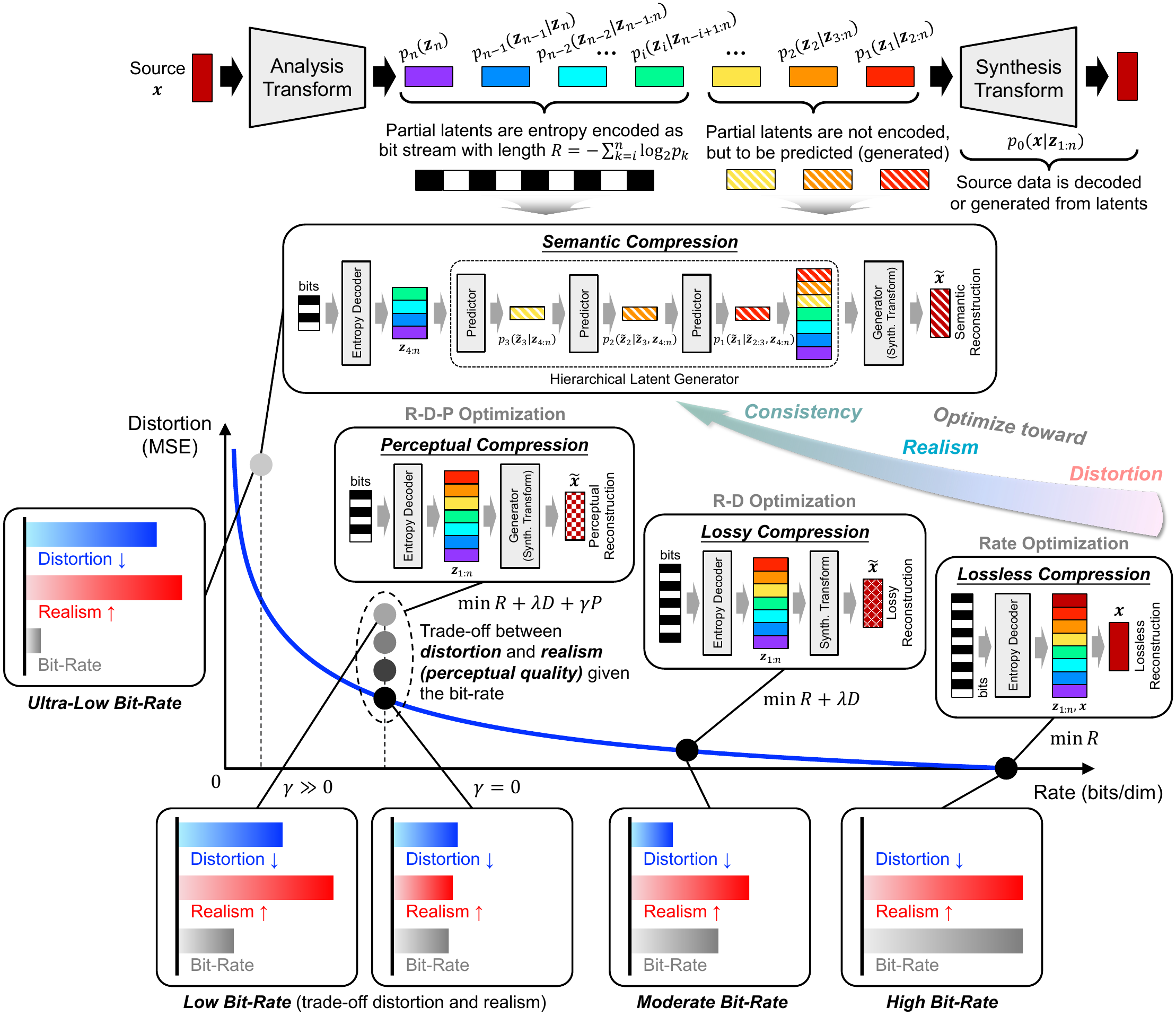}}
	\caption{Illustration of different compression paradigms enabled by deep generative models. In this figure, the solid color blocks stand for latent variables or source data that have been encoded as bit sequences to be transmitted or stored, and the dashed color blocks represent the ingredients that are not encoded but to be predicted using the probabilistic information provided by generative models. Here, $R$ denotes bit-rate cost, $D$ denotes distortion, and $P$ denotes the perceptual quality (realism) metric. The upward arrow ``$\uparrow$'' indicates ``higher value is better'', and vice versa. The change trends of distortion and realism with bit-rate are illustrated.}
	\label{Fig1}
\end{figure*}

Before we jump into neural compression with deep generative modeling, it is beneficial to remind ourselves what is data compression. In this article, as shown in Fig. \ref{Fig1}, we distinguish four types of data compression, namely
\begin{itemize}
	\item \emph{Lossless compression:} method that preserves all information and reconstructions are error-free, it holds for no distortion and the highest realism (best perceptual quality) while also being of the highest bit-rate cost (around 10 bits per pixel (bpp) for images);
	
	\item \emph{Lossy compression:} data is not preserved completely by a compression method, whose pursuit is a trade-off between the compression bit-rate and the reconstruction distortion (e.g., mean squared error (MSE)), i.e., the rate-distortion (R-D) trade-off \cite{balle2020nonlinear}. In this R-D optimization mode, the realism of reconstruction often shows clear degradation synchronized with the increase of distortion, and the bit-rate cost is typically $>$ 1 bpp for high quality;
	 
	\item \emph{Perceptual compression:} imperceptible high-frequency details are removed at encoding to lower bit rates, these details, however, are regenerated via generative models in the decoding phase, ensuring visually pleasing results, the whole compression system is optimized towards the rate-distortion-perception (R-D-P) trade-off \cite{blau2019rethinking}. From a single compressed representation, the decoder can decide to either reconstruct a low distortion reconstruction that is close to the input, a realistic reconstruction with high perceptual quality, or anything in between \cite{agustsson2023multi}. This mode typically works with the rate of 0.2 to 1 bpp;
	
	\item \emph{Semantic compression:} only an extremely small subset of latent representations are encoded as bit sequence for transmission or storage purpose, the system can then fully synthesis the uncoded latents with decoded counterparts and semantic labels as conditions. Operating at extremely low rates (typically below 0.02 bpp), this method relies heavily on the capability of generative models to produce realistic outcomes \cite{li2024misc}. The challenge lies in devising reasonable semantic guidance to ensure the reconstructed data remains \emph{consistent content} with the original counterpart. The optimization goal within semantic compression endeavors to balance \emph{realism} (also referred to as \emph{perceptual quality} in \cite{blau2019rethinking}) with reconstruction \emph{consistency}.
\end{itemize}

In designing a compression algorithm, the goal is to create a uniquely decodable code that minimizes the expected length of the encoded data, ideally approximating the data's entropy as closely as possible. The architecture of a typical compression system encompasses two primary components: an encoder and a decoder. This architecture is similar to that of a variational autoencoder (VAE) \cite{jakub2022deep}, a class of likelihood-based generative model, yet there are notable differences in their operational focus and objectives.

In terms of compression, the primary concern is the efficient transmission or storage of data as \emph{a compact bitstream}. Conversely, in the context of VAEs or general representation learning cases, the focus traditionally lies on the manipulation and interpretation of floating-point representations of latent codes, without particular constraints on the bitstream's length. From an information theory and source coding standpoint, the relationship between the dimensional attributes of a latent code and the length of its corresponding encoded bitstream is not inherently apparent. To bridge this gap and effectively transform a VAE into a practical compression scheme, additional steps are required to ensure efficient encoding latent representations into a bitstream suitable for storage or transmission.

In the encoding phase of data compression, the objective is to convert the source data into a discrete representation. These \emph{hierarchical latent representations} are not restricted to binary form; it can embody various discrete states. While the \emph{analysis transform} applied to the input data can be invertible, this is not an essential criterion for the process. An invertible transformation allows for the possibility of lossless compression, as the original data can be perfectly reconstructed using the inverse of the transformation in the decoding phase. However, in most scenarios where the transformation is not invertible, some information from the original data is inevitably lost, categorizing the process as lossy compression, depending on the nature and extent of information loss.

Following the analysis transform, discrete latent representations undergo a lossless encoding into a bitstream. This process involves mapping the discrete latent codes to a variable-length bitstream. A critical aspect in this stage is the use of entropy codecs, such as arithmetic codecs, which leverage knowledge about the probability of occurrence of each symbol in the latent representation. Hence, a pivotal factor in the success of this transform coding architecture is accurate probability estimation of the latent variables. Deep generative models play a significant role here, offering solutions for probability modeling in the latent space. The integration of deep generative modeling in this context not only enhances the efficiency of the encoding process but also significantly contributes to the overall effectiveness of data compression systems.

During the decoding phase of a compression system, the transmitted message -- represented as a bitstream -- is first received and then interpreted back into a discrete latent representation through the entropy decoder. Subsequent to the entropy decoding, an inverse transform (\emph{synthesis transform}) is applied to the quantized latents. It is important to note that this inverse transform does not necessarily have to be the exact inverse of the transform used in the encoding phase. As depicted in Fig. \ref{Fig1},  the choice of the inverse transform is dependent on the nature of encoding process and the desired distortion of reconstructed data. In the case of lossless compression, where the analysis transform is invertible, the synthesis transform can perfectly reconstruct the original data. But in lossy, perceptual, or semantic compression scenarios, where some information (latent variables) may have been discarded during encoding, the inverse transform aims to reconstruct an approximation of the original data, the prediction of uncoded latents and data is guided by the probability knowledge provided by generative models as in Fig. \ref{Fig1}.

Regarding the application of quantization schemes to these hierarchical latents, compression algorithms can be broadly classified into two categories. They differ in their approach to integrating deep generative models for quantifying the statistical properties of latent codes. As well-known, quantization method is also highly tied with the generation capability of various generative models. They are described as follows.

\subsection{Infinite Quantization with Continues Generative Modeling}

\begin{figure*}[t]
	\setlength{\abovecaptionskip}{0.cm}
	\setlength{\belowcaptionskip}{-0.cm}
	\centering{\includegraphics[scale=0.47]{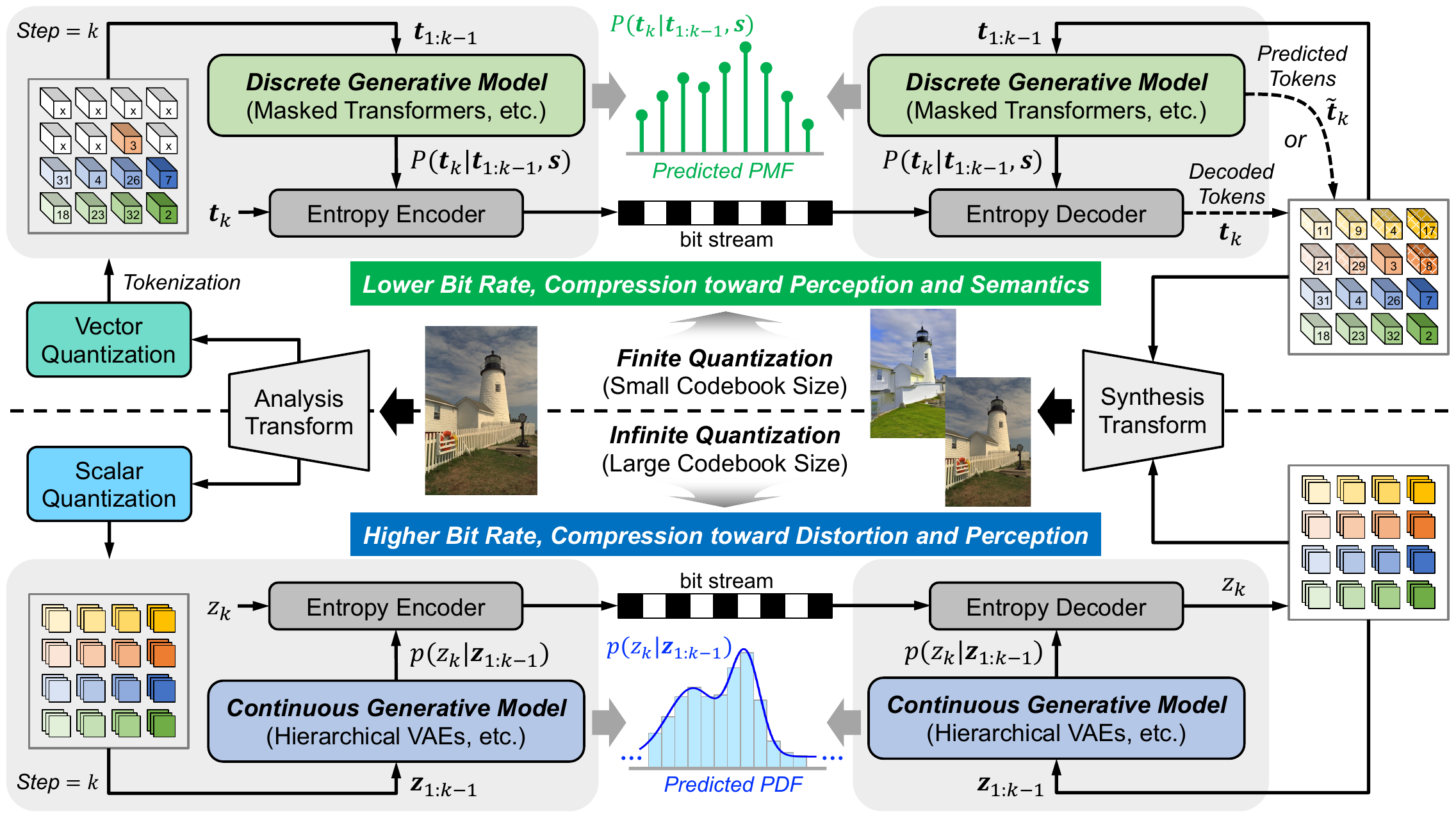}}
	\caption{Overview of data compression systems enabled by deep generative modeling. This figure presents two distinct roadmaps tailored for high and low bit-rate regions, respectively. It illustrates the integration of continuous and discrete generative models strategically employed to regulate the bit-rate of latent variables in data compression systems, $\bm{z}$ and $\bm{t}$ denote quantized latent codes, $\bm{s}$ denotes semantic guidance.}
	\label{Fig2}
\end{figure*}

As depicted in the lower section of Fig. \ref{Fig2}, following the application of an analysis transform -- commonly implemented using deep neural networks to exploit their nonlinear properties -- a scalar quantization (SQ) process is applied to each symbol within latent space. This \emph{infinite SQ} process, characterized by an effectively infinite codebook size, typically quantizes each latent symbol to its nearest integer by rounding. These scalar-quantized latent symbols are then encoded into a bitstream via entropy coding. Entropy codecs, such as arithmetic codecs, necessitate an estimate of the probability distribution over latent SQ symbols. Possessing this distribution estimate enables these codecs to compress the discrete latents into a bitstream losslessly. While an in-depth review of the arithmetic coding process is beyond the scope of this discussion, it is essential to recognize the significant role of continuous generative models in estimating the probability of latent symbols accurately.

An interesting aspect of modern arithmetic coding is its adaptive variant, which enables dynamic adjustment of symbol probabilities during the sequential compression process. This adaptive approach, also known as progressive coding, is particularly aligned with the capabilities of contextual learning in deep generative models. Notable examples of such models include auto-regressive models, hierarchical VAEs (e.g., top-down VAE and ResNet VAE) \cite{jakub2022deep}, etc., which capture the contextual dependency among latent variables. The interconnections between deep generative modeling and data compression, especially in terms of probability density function (PDF) estimation (also known as entropy modeling), showcases how advancements in generative AI models can be integrated with data compression methods to achieve improved performance and new capabilities in data handling and storage.

In the context of infinite SQ, probabilistic generative models are crucial in managing the compressed bit-rate. With unbounded SQ, where the range of integers extends beyond the limitations imposed by the analysis transform, control over the bit-rate is constrained through the entropy model applied to the latents. This approach (referred to as nonlinear transform coding (NTC) \cite{balle2020nonlinear}), integrating infinite SQ with generative modeling, is particularly relevant to \emph{high bit-rate} scenarios as depicted in Fig. \ref{Fig1} and Fig. \ref{Fig2}. These scenarios include lossless, lossy, and certain high bit-rate perceptual compression cases. The flexibility of this method allows for tailored optimization of R-D or R-D-P trade-offs. Given the bit-rate constraint, the optimization goal predominantly concentrates on minimizing the distortion of the reconstructed data in this high rate region.

\subsection{Finite Quantization with Discrete Generative Modeling}

\begin{figure*}[t]
	\setlength{\abovecaptionskip}{0.cm}
	\setlength{\belowcaptionskip}{-0.cm}
	\centering{\includegraphics[scale=0.366]{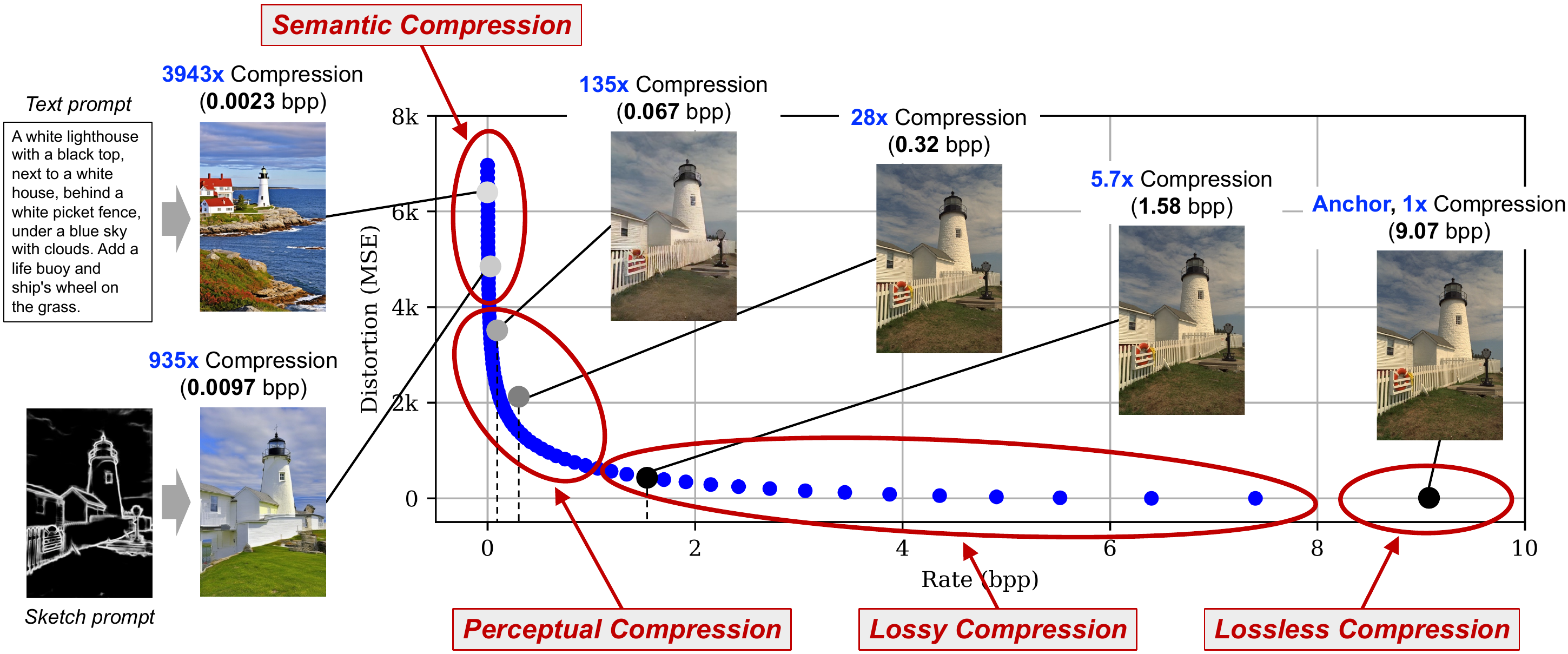}}
	\caption{Illustrating four types of image compression in terms of the rate-distortion curve. Lossless compression, targeting pixel-perfect reconstructions, typically results in higher bit rates (around 9.07 bits per pixel (bpp)) compared to lossy compression. Neural lossy compression algorithms generally operate above 0.1 bpp, with distortions becoming nearly imperceptible above 1 bpp, as visually evidenced. However, at rates below 1 bpp, lossy compression can introduce blurry artifacts, impacting visual perception. Perceptual compression, operating between 0.02 and 1 bpp, is designed to generate realistic textures, maintaining perceptual quality at reduced rates, as seen in third image from the right. When rates fall below 0.1 bpp, the syntactic fidelity drops, and textures become distorted. At extreme low bit-rates (below 0.02 bpp), where traditional methods falter, semantic compression steps in, focusing on preserving essential visual information. Utilizing advanced generative models, semantic compression can generate meaningful reconstructions from minimal data, such as a short text prompt (0.0023 bpp) or a sketch prompt (0.0097 bpp), preserving key semantic elements and structural information.}
	\label{Fig3}
\end{figure*}

In low bit-rate regions, typically associated with perceptual and semantic compression, the encoding process also involves transforming input data, such as an image, into a latent representation, often a sequence of feature vectors. This is shown in the upper section of Fig. \ref{Fig2}. Each feature vector undergoes \emph{finite vector quantization (VQ)}, where it is matched to the closest vector in a learned codebook, an operation known as tokenization. Consequently, the source data is represented as a sequence of discrete tokens. Due to the finite size of a VQ codebook, a natural upper limit is imposed on the bit-rate cost when these token indices are converted into bits. By integrating discrete generative models, the contextual relationships among these tokens are elucidated (modeling the probability mass function (PMF) of each token) \cite{el2022image}, further reducing the bit-rate via entropy coding. This also enables the representation of tokens in a variable-length bitstream format. Furthermore, the generative models' robust generation capabilities facilitate ultra-low bit-rate applications. In this region, given the bit-rate constraint, the dominant optimization goal shifts as improving the realism of the reconstructed data and its consistency with the original one, i.e., balancing realism with consistency.

In tokenization-based compression, the potential of \emph{pre-trained masked Transformers (MT)}, so-called \emph{foundation models}, can be effectively leveraged for generating content across various modalities \cite{deletang2023language}. Accordingly, discrete tokens, analogous to lexical words, are input into MT for contextual generative modeling. It is important to recognize that learned VQ and its modern variants continue to be the dominant techniques for transforming continuous hierarchical latents into discrete tokens, particularly in environments requiring very low bit-rate compression. In such scenarios, VQ effectively exploits correlations among latent dimensions. However, with increasing bit-rate demands, optimizing VQ presents challenges, leading to the well-documented issue of underutilized VQ codebooks \cite{mentzer2023finite}. As the codebook size expands, many codewords remain unused, diminishing their utility. Consequently, in high bit-rate scenarios, SQ often becomes the preferable method due to its ability to manage larger codebooks more efficiently.

Fig. \ref{Fig3} comprehensively presents a rate-distortion curve including four types of compression -- lossless, lossy, perceptual, and semantic -- alongside illustrative reconstructions for each. The blue numbers mark the relative compression ratio versus the anchor scheme of lossless compression.

\section{Deep Generative Modeling for Transmission: The Functionality of Resiliency}\label{section_transmission}

\begin{figure*}[t]
	\setlength{\abovecaptionskip}{0.cm}
	\setlength{\belowcaptionskip}{-0.cm}
	\centering{\includegraphics[scale=0.41]{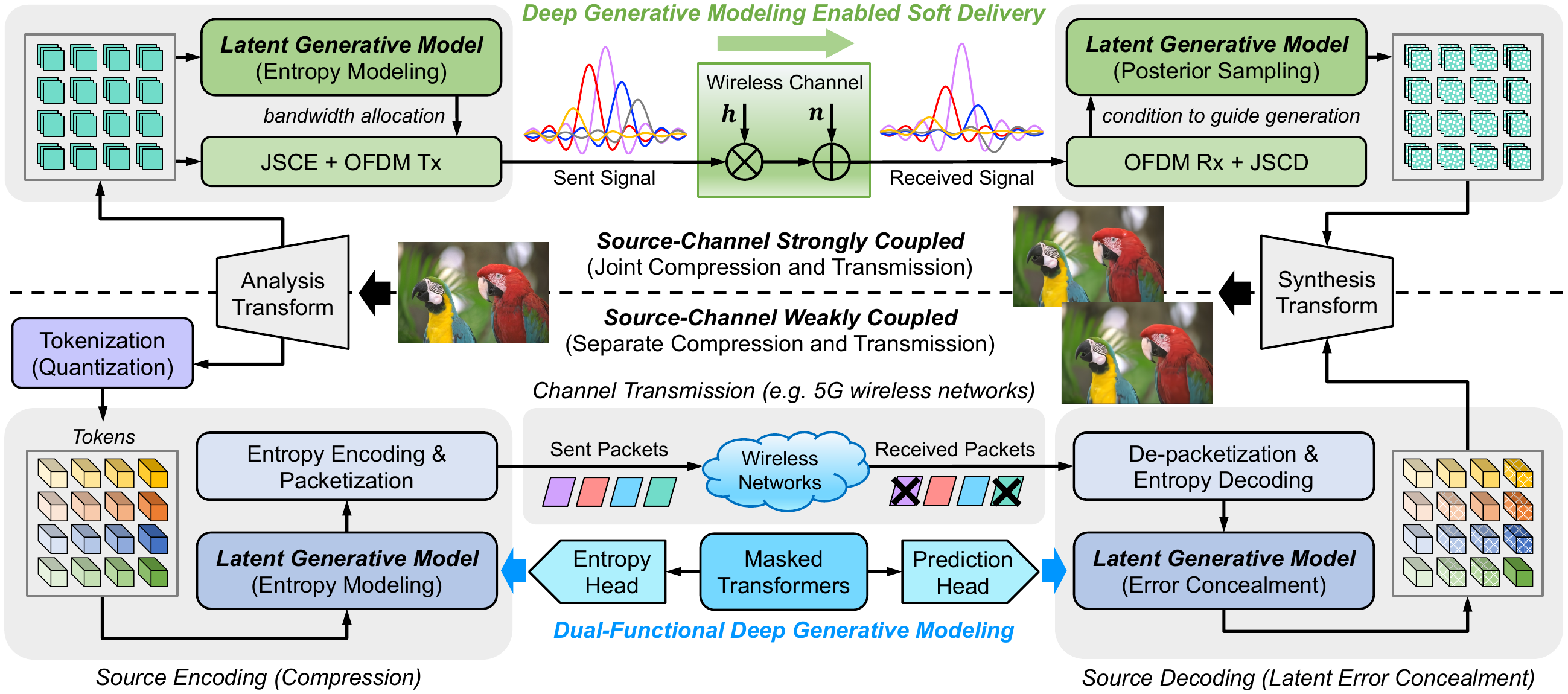}}
	\caption{Overview of data transmission systems enabled by deep generative models. This figure presents two distinct roadmaps tailored for source compression and channel transmission strongly and weakly coupled, respectively.}
	\label{Fig4}
\end{figure*}

The rise of immersive technologies such as volumetric video, virtual reality, and video conferencing is driving increased demands on end-to-end communication systems. This trend challenges the traditionally separate development paths of source compression and channel transmission domains. In the face of very big data transmission and massive emerging applications, on the one hand, current wireless transmission systems suffer from time-varying channel conditions, in which case the mismatch between communication rate and channel capacity results in obvious \emph{cliff-effect}, i.e., the performance breaks down when the channel capacity goes below communication rate. On the other hand, the widely-used entropy coding in compression, is quite sensitive to the channel decoded results, any left bit error can lead to catastrophic error propagation in entropy decoding. It is thus the very time to develop new \emph{resilient} or \emph{robust} coded transmission paradigms to support real-time communications (RTC) toward emerging media end-to-end transmission tasks.

Under this background, deep generative modeling, inspired by generative compression concepts, also offers innovative solutions to formulate joint source and channel coding (JSCC) systems, addressing the challenges of end-to-end communication. Unlike traditional transmission systems that emphasize instance matching between transmitter and receiver, generative modeling focuses on optimizing the distribution match between transmitted and recovered data. This approach paves the way for various types of learned JSCC systems.

One extreme case within this JSCC paradigm involves a strongly-coupled system, where compression and transmission functionalities are integrated into a single module established with deep neural networks. Here, the JSCC module learns to directly maps source data to a modulated waveform, similar to amplitude-modulated analog communication systems. In such systems, likelihood-based VAEs are instrumental in formulating and optimizing the process end-to-end. Conversely, given the prevailing modularity and separation-based digital design in contemporary communication systems, we also propose a novel weakly-coupled JSCC system as an alternative approach. This system emphasizes latent space packet loss concealment, predicting by utilizing the strengths of deep generative models. Such a weakly-coupled system offers a compatible approach to realize JSCC, thus can plug-and-play with today's standardized communication systems. We discuss them in detail.

\subsection{Strongly-Coupled JSCC: End-to-End Robust Transmission}

\begin{figure*}[t]
	\setlength{\abovecaptionskip}{0.cm}
	\setlength{\belowcaptionskip}{-0.cm}
	\centering{\includegraphics[scale=0.65]{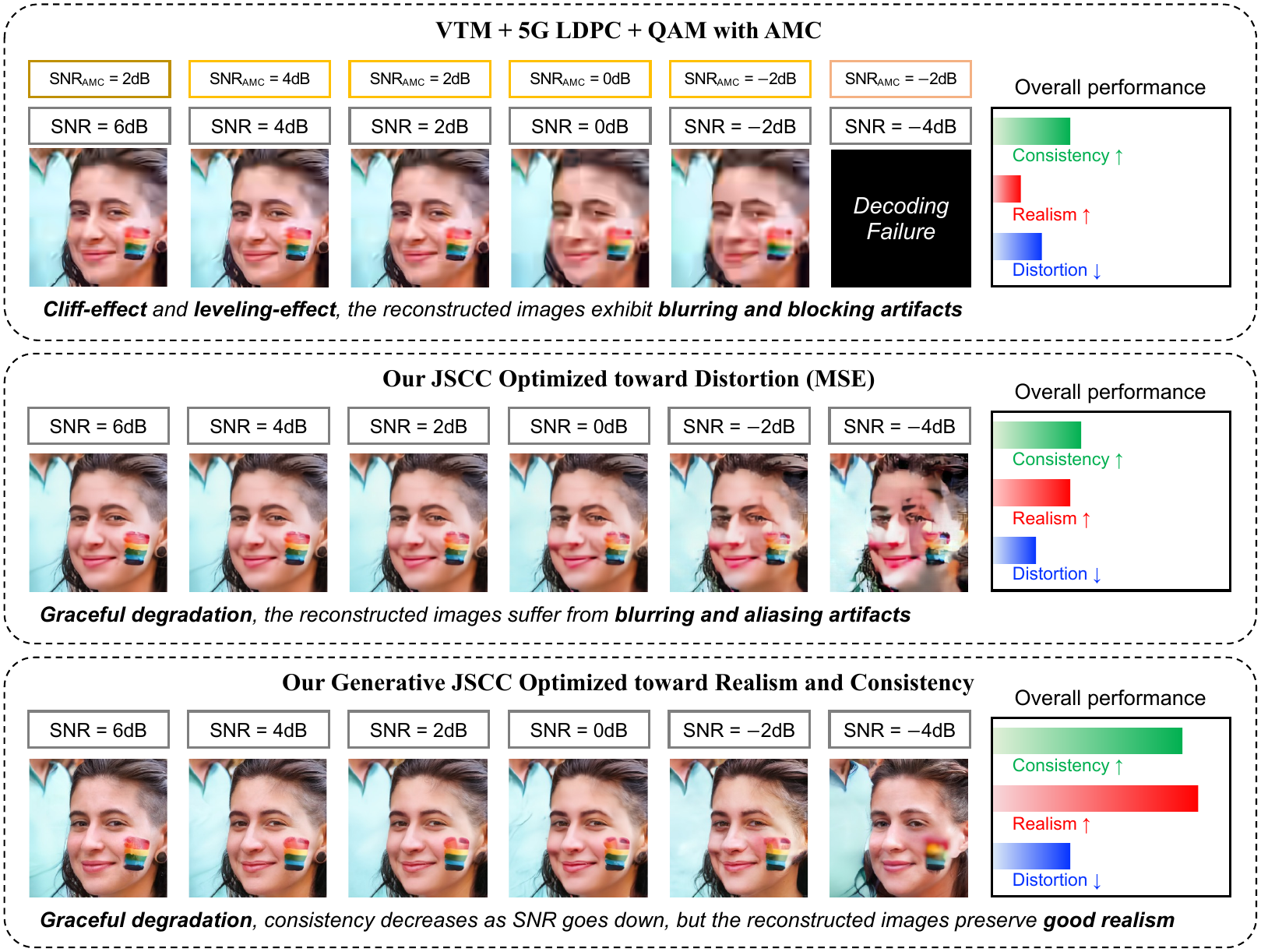}}
	\caption{Performance comparison of three state-of-the-art wireless image transmission methods, where the    example photo is sampled from the widely used FFHQ dataset (specifically image ID: 69037). The upward arrow ``$\uparrow$'' indicates ``higher value is better'', and vice versa. For each method, we visualize their reconstructions with decreasing channel signal-to-noise ratio (SNR) from 6dB to $-$4dB. The visual quality of these methods is evaluated in terms of consistency, realism, and distortion, measured by LPIPS, FID, and MSE metrics, respectively. In particular, for VTM + 5G LDPC + QAM with the AMC mechanism, $\text{SNR}_{\text{AMC}}$ in the first row denotes the estimated SNR used to select the coding rate of LDPC code and the QAM level, corresponding to the actual channel SNR displayed in the second row. }
	\label{Fig5}
\end{figure*}

A pioneering exploration into utilizing deep learning for strongly-coupled joint source and channel coding (JSCC) led to the development of what is known as Deep JSCC \cite{bourtsoulatze2019deep}. This approach employs an autoencoder architecture that is trained end-to-end, with the wireless channel status integrated as an additional input at the bottleneck layer. The encoder in this system is designed to learn a direct mapping from the source data, such as images, to a set of channel input symbols, with the decoder performing the inverse operation. This end-to-end learning approach facilitates the use of loss measures between the input and output of the autoencoder.

Following initial advancements, subsequent studies have revealed that purely distortion-optimized Deep JSCC systems, despite their robust performance and absence of a cliff-effect, do not always outperform advanced separation-based source and channel coding schemes (such as VTM + 5G LDPC for image transmission) especially for large-size source data (e.g., high-resolution images). This limitation is attributed to the inherent nature of fixed-length JSCC and the \emph{discriminative modeling} approach of autoencoders.

A significant breakthrough in this domain was the development of nonlinear transform source-channel coding (NTSCC) \cite{dai2022nonlinear}, which introduced a variable-length JSCC system learned through \emph{generative modeling} as depicted in the upper section of Fig. \ref{Fig4}. The NTSCC framework, drawing inspiration from R-D optimized VAEs analogous to those utilized in NTC for data compression, concentrates on optimizing the trade-off with channel bandwidth cost and end-to-end distortion. This methodology ensures optimal bandwidth usage while minimizing distortion impacts. Additionally, the integration of a robust latent generator, such as a latent diffusion model, facilitates posterior sampling. This sampling is guided by either the raw received signal or the JSCC decoded outcome, serving as a refined conditioning mechanism. The result is the generation of realistic and consistent reconstructions. These advancements in strongly-coupled JSCC underscore the significant impact of generative modeling in enhancing system performance.

%

To intuitively demonstrate the effect of the deep generative prior applied to JSCC, we conduct a wireless image transmission experiment with the results shown in Fig. \ref{Fig5}. We evaluate three representative coded transmission approaches for comparison, including (i) the state-of-the-art separation-based transmission scheme VTM + 5G LDPC + QAM with adaptive coding and modulation (AMC) mechanism; (ii) the strongly-coupled JSCC \cite{wang2023improved}; and (iii) our generative JSCC transmission scheme that exploits the received corrupted JSCC signal as the fine-grained condition to guide posterior sampling with a pre-trained powerful generator. For the separation scheme, the LDPC code length is 4096. The specific modulation order and channel coding rate are selected according the MCS Table in 3GPP TS 38.214 standard. For a fair comparison, we constraint the bandwidth ratio (defined as the number of transmitted OFDM resource elements divided by the number of source data dimensions) for all schemes to no more than 0.02, and select decreasing channel SNRs from 6dB to $-$4dB. Note that this configuration is challenging due to the combination of very low bandwidth cost and severe channel conditions.

Our main observations from results in Fig. \ref{Fig5} are as follows:
\begin{itemize}
	\item VTM + 5G LDPC performs well under accurate estimated channel conditions. 
	However, when the channel SNR drops below the estimated SNR used in AMC, it could fail to reconstruct transmitted image due to the inadequate protection from channel coding (cliff-effect). 
	Conversely, when channel SNR goes beyond estimated SNR, it does not yet improve the quality of reconstruction (leveling-effect).
	Our JSCC schemes, in contrast, is demonstrated to be robust against channel quality variations and exhibits graceful degradation as the channel SNR goes down.
	
	\item VTM + 5G LDPC and MSE-optimized JSCC schemes are primarily designed to minimizing signal-level distortions (as measured by PSNR). 
	As channel SNR decreases, their reconstructions become less realistic and present various artifacts.
	In specific, the VTM + 5G LDPC exhibits the blurring and blocking artifacts, while MSE optimized JSCC scheme suffers from aliasing artifacts. 
	These noisy artifacts degrade the realism of their reconstructions.
	
	\item Our generative JSCC can provide realistic and visually pleasing reconstructions even in regions of extremely-low SNR. 
	It significantly differs from the other two schemes in that it can maintain perfect realism, with a consistency decrease in image content as the SNR decreases.
\end{itemize}

\subsection{Weakly-Coupled JSCC: Packet Loss-Resilient Transmission}

\begin{figure*}[t]
	\setlength{\abovecaptionskip}{0.cm}
	\setlength{\belowcaptionskip}{-0.cm}
	\centering{\includegraphics[scale=0.7]{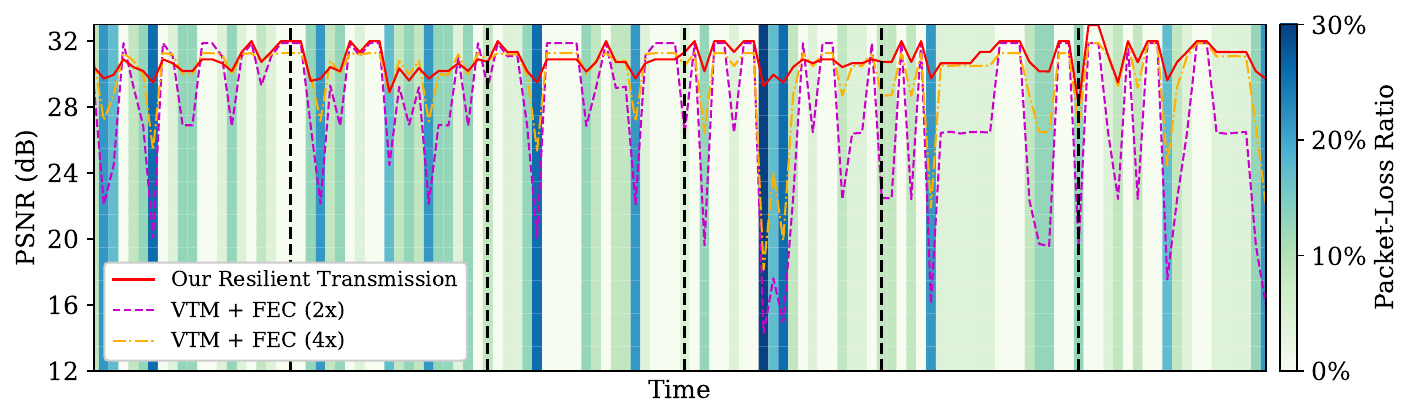}}
	\caption{Comparison of multiple resilient image transmission schemes on the Kodak dataset over a network with variable real-time packet loss. The black dashed line represents the computation of the average packet loss rate from the previous interval, which informs the choice of coding methods for subsequent intervals. Specifically, interpacket FEC codes are employed to preemptively add redundant packets -- each configuration should add $r$ redundancy packets to compensate for up to $r$ expected lost packets prior to transmission. The resilience against packet loss is assessed using the VTM-20.0 codec enhanced with varying levels of FEC, denoted as ``VTM + FEC ($N\times$)'', where $N$ signifies the multiplication factor relative to the average packet loss.}
	\label{Fig6}
\end{figure*}

To ensure compatibility with existing digital communication systems, we explore the advantages of incorporating deep generative modeling into weakly-coupled JSCC design. This approach maintains the traditional elements of channel coding and digital modulation found in current systems, while adapting the source compression component to accommodate the challenges of unreliable wireless networks, particularly those prone to packet loss. In traditional scenarios, compensating for packet loss typically involves re-transmissions. However, this method proves inefficient for applications necessitating real-time visual interactions such as online meetings, extended reality (XR) applications, and cloud gaming due to its inherent delays. These applications demand strategies that ensure minimal latency. To manage lost packets within the strict time constraint, the concept of weakly-coupled JSCC will become pivotal. This approach optimizes transmission efficiency and enhances resiliency against burst packet losses.

Drawing inspiration from the remarkable generative capability of masked Transformers (MT), this joint optimization approach combines entropy modeling and latent packet-loss concealment within a unified framework focused on modeling the contextual relationships in latent space. As illustrated in the bottom panel of Fig. \ref{Fig4}, MT-based latent generative models (e.g., bi-directional Transformers in \cite{chang2022maskgit}) are exploited to understand and leverage the \emph{casual-order contextual dependencies} among latents. This model performs two critical functionalities within a single model thus facilitating practical deployment: it estimates conditional probability distributions over quantized tokens, enhancing compression efficiency, and it predicts and replaces missing latent values, improving resilience against data loss. By optimizing the system’s ability to handle errors alongside its efficiency, this approach significantly enhances the overall performance, making it well-suited for the demanding requirements of real-time applications.

Fig. \ref{Fig6} evaluates various resilient transmission strategies over a lossy network, utilizing a Markov chain-based packet-loss trace. These traces are widely-recognized for effectively simulating the burst-like packet loss characteristics in real-time WLAN or mobile networks. For all schemes, the network's packet loss status is continuously monitored, and the average packet loss ratio over a historical interval is computed to inform the selection of coding strategies for subsequent intervals. This includes the adaptation of coding modes in our resilient transmission scheme or the integration of forward error correction (FEC) codes in traditional codec.

Our main observations from Fig. \ref{Fig6} are as follows. In widely-used separate coding schemes such as VTM combined with FEC, the exact packet loss ratio cannot be predetermined; only a predictive loss rate is available. Our investigations into various FEC redundancy levels reveal that excessive redundancy compromises reconstruction quality, while insufficient protection leads to decoding failures. Particularly in severe scenarios where packet loss rates range from 20\% to 30\%, even quadrupling the FEC does not prevent decoding failures. In contrast, our proposed dual-functional generative model, integrated within a weakly-coupled JSCC system, demonstrates superior resilience. This system dynamically manages quality variations during real-time packet loss, showing notable improvements in both decoding reliability and visual quality compared to the traditional VTM + FEC methods. Additionally, our approach demonstrates excellent compression efficiency in the absence of packet loss, confirming that our scheme effectively balances efficiency and resilience through the strategic use of generative modeling.

\section{Conclusion and Discussion}

In this article, we have explored the pivotal role of generative AI in advancing source compression and end-to-end transmission techniques, illustrating how generative modeling significantly enhances performance in these areas. 

However, it is worth considering the reverse perspective: in modern generative AI systems, including GPT-4, compression or probabilistic modeling seems to be a fundamental component. This leads to an interesting question: Can compression idea alone pave the way to general intelligence or even consciousness? Compression facilitates learning data correlations and distributions, but certain aspects of intelligence, such as causal reasoning, logical thinking, or abstract conceptualization, out of mere data patterns. This observation suggests that although advancements in compression and transmission underscore the significant strides made in generative AI, they might not be the sole avenues to achieving higher forms of intelligence. Future advancements in generative AI may require integrating these capabilities with more sophisticated cognitive functions to truly mirror the complexities of human intelligence. Such integration has the potential to push compression and transmission techniques into a new era marked by exceptional intelligence and conciseness.

\bibliographystyle{IEEEbib}
\bibliography{myRef}

\end{document}